\documentclass[aps,prl,reprint,superscriptaddress,nofootinbib]{revtex4-2}
\usepackage{blindtext}
\usepackage{lipsum}
\usepackage{graphics}
\usepackage{amsmath}
\usepackage{graphicx}
\usepackage{graphics}
\usepackage{amssymb}
\usepackage{verbatim}
\usepackage{physics}
\usepackage{verbatim}
\usepackage{float}
\usepackage{dsfont}
\usepackage[dvipsnames]{xcolor}
\usepackage{framed}
\definecolor{shadecolor}{RGB}{224,224,224}

\makeatletter
\def\l@subsubsection#1#2{}
\makeatother

\usepackage[dvipsnames]{xcolor}
\usepackage[normalem]{ulem}
\usepackage{wasysym} 
\usepackage[export]{adjustbox}

\makeatletter
\DeclareFontFamily{OMX}{MnSymbolE}{}
\DeclareSymbolFont{MnLargeSymbols}{OMX}{MnSymbolE}{m}{n}
\SetSymbolFont{MnLargeSymbols}{bold}{OMX}{MnSymbolE}{b}{n}
\DeclareFontShape{OMX}{MnSymbolE}{m}{n}{
    <-6>  MnSymbolE5
   <6-7>  MnSymbolE6
   <7-8>  MnSymbolE7
   <8-9>  MnSymbolE8
   <9-10> MnSymbolE9
  <10-12> MnSymbolE10
  <12->   MnSymbolE12
}{}
\DeclareFontShape{OMX}{MnSymbolE}{b}{n}{
    <-6>  MnSymbolE-Bold5
   <6-7>  MnSymbolE-Bold6
   <7-8>  MnSymbolE-Bold7
   <8-9>  MnSymbolE-Bold8
   <9-10> MnSymbolE-Bold9
  <10-12> MnSymbolE-Bold10
  <12->   MnSymbolE-Bold12
}{}

\let\llangle\@undefined
\let\rrangle\@undefined
\DeclareMathDelimiter{\llangle}{\mathopen}%
                     {MnLargeSymbols}{'164}{MnLargeSymbols}{'164}
\DeclareMathDelimiter{\rrangle}{\mathclose}%
                     {MnLargeSymbols}{'171}{MnLargeSymbols}{'171}
\makeatother

\usepackage{tikz}
\usepackage{tikz-cd}
\usetikzlibrary{arrows}
\usetikzlibrary{snakes}
\usetikzlibrary{intersections}
\usetikzlibrary{shapes.geometric}
\usetikzlibrary{decorations.pathmorphing, patterns,shapes}
\usetikzlibrary{decorations.markings}

\tikzset{
	partial ellipse/.style args={#1:#2:#3}{
		insert path={+ (#1:#3) arc (#1:#2:#3)}
	}
}

\tikzset{
	mid arrow/.style={postaction={decorate,decoration={
				markings,
				mark=at position .575 with {\arrow[#1]{stealth}}
	}}},
	near arrow/.style={postaction={decorate,decoration={
				markings,
				mark=at position .275 with {\arrow[#1]{stealth}}
	}}},
	far arrow/.style={postaction={decorate,decoration={
				markings,
				mark=at position .800 with {\arrow[#1]{stealth}}
	}}},
}

\pgfdeclarepatternformonly{south west lines}{\pgfqpoint{-0pt}{-0pt}}{\pgfqpoint{3pt}{3pt}}{\pgfqpoint{3pt}{3pt}}{
	\pgfsetlinewidth{0.4pt}
	\pgfpathmoveto{\pgfqpoint{0pt}{0pt}}
	\pgfpathlineto{\pgfqpoint{3pt}{3pt}}
	\pgfpathmoveto{\pgfqpoint{2.8pt}{-.2pt}}
	\pgfpathlineto{\pgfqpoint{3.2pt}{.2pt}}
	\pgfpathmoveto{\pgfqpoint{-.2pt}{2.8pt}}
	\pgfpathlineto{\pgfqpoint{.2pt}{3.2pt}}
	\pgfusepath{stroke}}

\definecolor{orange(ryb)}{HTML}{FFA500}
\definecolor{lightorange(ryb)}{HTML}{FFB300}
\definecolor{dodgerblue}{HTML}{1E90FF}
\definecolor{lightdodgerblue}{HTML}{4dbff7}
\definecolor{crimson}{HTML}{FF4C4C}
\definecolor{pinkerton}{HTML}{EC368D}
\definecolor{forest}{HTML}{6DD189}
\definecolor{lightishgray}{HTML}{DFDFDF}
\definecolor{error-red}{HTML}{EFB2B6}

\usepackage[colorlinks=true,citecolor=dodgerblue,linkcolor=dodgerblue,urlcolor=dodgerblue,pdftitle={Gluing TNS}]{hyperref}

\begin{document}

\title{Finite-Depth Preparation of Tensor Network States from Measurement}
\author{Rahul Sahay}
\affiliation{Department of Physics, Harvard University, Cambridge, Massachusetts 02138 USA}
\author{Ruben Verresen}
\affiliation{Department of Physics, Harvard University, Cambridge, Massachusetts 02138 USA}
\affiliation{Department of Physics, Massachusetts Institute of Technology, Cambridge, MA 02139, USA}

\date{\today}

\begin{abstract}
Although tensor network states constitute a broad range of exotic quantum states, their realization is challenging and often requires resources whose depth scales with system size.
In this work, we explore criteria on the local tensors for enabling deterministic state preparation via a single round of measurements and on-site unitary feedback.
We use these criteria to construct families of measurement-preparable states in one and two dimensions, tuning between distinct symmetry-breaking, symmetry-protected, and intrinsic topological phases of matter.
For instance, in one dimension we chart out a three-parameter family of preparable states which interpolate between the AKLT, cluster, GHZ and other states of interest.
Our protocol even allows one to engineer preparable quantum states with a range of desired correlation lengths and entanglement properties.
In addition to such constructive approaches, we present diagnostics for verifying whether a given tensor network state is preparable using measurements.
We conclude by charting out generalizations, such as considering multiple rounds of measurements, implementing matrix product operators, and using incomplete basis measurements.
\end{abstract}

\maketitle

\textbf{Introduction.} In recent years, projective measurements and unitary feedback have emerged as powerful tools in the quest to realize interesting states of matter in quantum devices \cite{Briegel01,Raussendorf05,Aguado08,Bolt16,piroli2021locc,LREfromSPT,verresen2021efficiently,lu2022shortcut,Bravyi22,lee2022decoding,hierarchy,zhu2023nishimori,smith2023aklt,shortestroute,gunn2023phases,li2023set,lu2023mixedstateLRO,piroli2024approximate,malz2024logdepth, li2023measuring, Wu_2023, sukeno2024LGT,buhrman2023state, lee2024symmetry}.
Explorations into the capabilities of these tools have led to both an increased theoretical understanding into the complexity of quantum states but also experimental advances in creating previously unrealized phases of matter \cite{iqbal2023topological,foss2023advantage,chen2023nishimori,Iqbal2024nonabel,Bluvstein_2023}.
Nevertheless, much remains to be understood regarding the landscape of quantum states that are deterministically preparable in this way.
This is especially the case for the realization of quantum states away from clean renormalization group fixed points. 
Indeed, while isolated deterministic examples \cite{smith2023aklt,zhu2023nishimori} have been discovered for one-dimensional quantum systems, organizing principles, broader classes of preparable quantum states, and higher-dimensional generalizations are lacking.

In this work, we study the preparation of tensor network states\footnote{Specifically, we consider those that fall within the general class of ``projected entangled pair states'' including the matrix product state and its two-dimensional generalizations.} \cite{Fannes1992,Cirac2021MPSReview, PhysRevLett.96.220601} which are known to capture a rich landscape of many-body quantum states. In particular, we will start from disentangled clusters of qudits and perform a single round of parallel entangling measurements, followed by on-site unitary feedback to guarantee a deterministic result.
Our approach proceeds by identifying a criteria on local tensors that ensures preparability, inspired in part by an initial work by Smith et al.~\cite{smith2023aklt}, and motivated by results presented in our companion paper \cite{sahay2024finite} that establishes this criteria as necessary and sufficient in a more restricted setting.
We will use this to (i) construct landscapes of preparable quantum states in one and two dimensions, (ii) develop settings where correlation and entanglement properties of a desired quantum state can be engineered, and finally (iii) develop diagnostics for determining when a many-body quantum state is preparable.

\textbf{Preparability condition.} Let $\ket{\Psi}$ be a tensor network state generated by a local tensor $A$. While our discussion applies to general dimensions, we sketch it in two dimensions:
\begin{equation} \label{eq:Psi}
\ket{\Psi} = \includegraphics[valign = c, scale = 0.5]{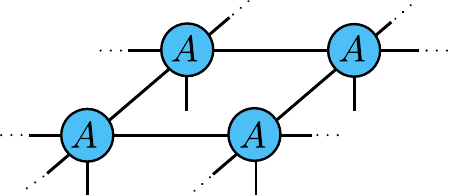}
\end{equation}
As usual, here we contract the `virtual' legs of $A$ (horizontal; dimension $\chi$), giving rise to $\ket{\Psi}$ on the uncontracted `physical' legs (vertical; dimension $d$).
The goal will be to prepare this state in finite time using measurement and feedback. The general methodology will be to start with decoupled clusters of qudits, each of which has a wavefunction given by a single $A$ tensor. Note that in this approach, all legs of $A$ correspond to actual qudits in the system; the `virtual' legs simply correspond to ancillas:
\begin{equation} \label{eq:cluster}
\ket{A}^{\otimes N} = \includegraphics[valign = c, scale = 0.5]{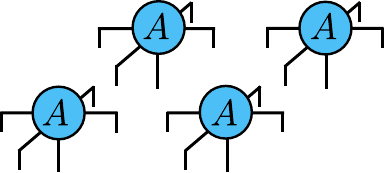}
\end{equation}

We now want to `fuse' these legs together. This is achieved by performing (generalized) Bell measurements on the ancilla qudits. Indeed, in the case of the ideal measurement outcome $\ket{\mathds{1}} = \frac{1}{\sqrt{\chi}} \sum_{a=1}^\chi \ket{a}\ket{a}$, we see that the post-measurement state is Eq.~\eqref{eq:Psi}. For more generic outcomes $\alpha \in \{1,2,\cdots,\chi^2\}$, however, we obtain the insertion of an operator $V_\alpha$ in the virtual bond:
\begin{equation}
\includegraphics[valign = c, scale = 0.6]{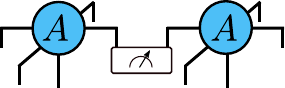}\ \  \longrightarrow\ \  \includegraphics[valign = c, scale = 0.6]{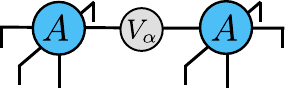}
\end{equation}
where $V_\alpha$ is the operator whose Choi state defines the measurement outcome $\ket{V_\alpha} = \frac{1}{\sqrt{\chi}} \sum_{a,b=1}^\chi \left( V_\alpha \right)_{a,b} \ket{a}\ket{b}$. 
In this work, we will consider the case where $V_\alpha$ is unitary, which is equivalent to the measurement basis $\{ \ket{V_\alpha} \}$ being maximally-entangled.
In fact, in our companion work \cite{sahay2024framework} we derive this as a necessary property for preparability in a broad range of circumstances. The resulting set of unitary operators $\{ V_\alpha \}$ then defines a trace-orthogonal basis of $\chi\times \chi$ matrices, also called a \emph{unitary error basis} \cite{knill1996group, klappenecker2005monomiality, klappenecker2002beyond}. Note that the ideal measurement outcome corresponds to $V_1 = \mathds{1}$.

In summary, in this approach we arrive at the desired state Eq.~\eqref{eq:Psi} with virtual operators inserted. Although these depend on the measurement outcome, deterministic state preparation requires that for each set of outcomes, there exists a unitary which can correct for them. I.e., we need\footnote{Here we have chosen an on-site unitary feedback. To the best of our knowledge this pertains to all known examples of deterministic measurement-based state preparation.}:
\begin{equation}
\includegraphics[valign = c, scale = 0.5]{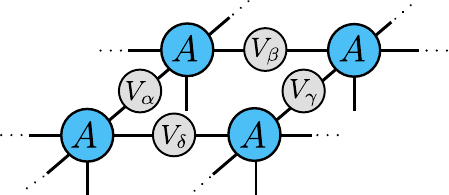}\! \! =\! \!  \includegraphics[valign = c, scale = 0.5]{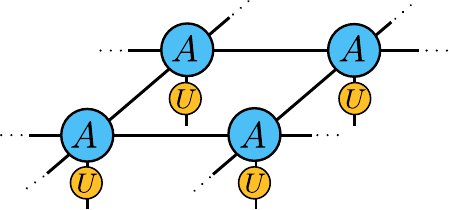}
\end{equation}
(with site- and measurement-dependent $U$'s in general) such that we can correct for the measurement outcome by applying $U^\dagger$ on the physical qubits. 
This daunting global condition becomes more weildy if we look for a local condition where a each $V_\alpha$ can be pushed through the tensor network:
\begin{equation}\label{eq:VA=AU}
\includegraphics[valign = c,scale = 0.65]{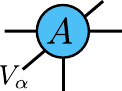} \; = \;  \includegraphics[valign = c, scale = 0.65]{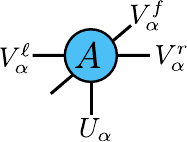}  
\end{equation}
This in turn requires the newly spawned virtual operators $V_\alpha^{l,r,f}$ to appropriately push through.

This rephrases the problem of measurement-based state preparation as a game of locally pushing operators until all measurement outcomes are reduced to unitaries on the physical degrees of freedom, where they can easily be corrected. This is an interesting many-body problem of sorts, and a full solution and characterization of tensors with these properties poses an exciting research question. In our companion paper, we show how this problem becomes tractable in at least one scenario---a one-dimensional setting where the flow of classical information is unidirectional. In the present work, we instead take a more phenomenological approach, demonstrating the power of this approach in a more general setting where a full solution is not yet available.

Before transitioning to concrete examples, we highlight that Eq.~\eqref{eq:VA=AU} can be proven to be \emph{equivalent} to the following symmetry condition on the doubled PEPS tensor:
\begin{equation} \label{eq:VTV=T}
\mathbb E = \includegraphics[valign = c, scale = 0.6]{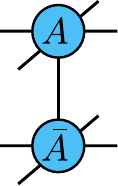} = \includegraphics[valign = c, scale = 0.6]{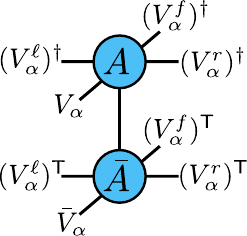}  
\end{equation}
where we introduced $\mathbb E$ following Ref.~\onlinecite{Schuch13}.
It is clear that Eq.~\eqref{eq:VA=AU} implies Eq.~\eqref{eq:VTV=T}, and while the other direction can be proven using Kraus' theorem, we provide a constructive proof for $U$ in the 
Supplemental Materials. 

\textbf{Landscape of Preparable $\chi=2$ MPS.} Let us explore how already the simplest instance of the above provides a remarkably rich phase diagram. We consider one-dimensional matrix product states (MPS) with bond dimension $\chi=2$. In this case, the only unitary error basis possible is given by the Pauli matrices $\{\mathds{1},X,Y,Z\}$ (up to global conjugation if $V_1 =\mathds{1}$) \cite{klappenecker2005monomiality}. If we now consider the one-dimensional analog of Eq.~\eqref{eq:VTV=T}, the left-hand side is the MPS transfer matrix with a natural solution:
\begin{equation}
 \mathbb E = \lambda_1 \mathds{1} \otimes \mathds{1} + \lambda_x X \otimes X + \lambda_y Y \otimes Y + \lambda_z Z \otimes Z,
\end{equation}
where the $\lambda_i$'s are required to be real, non-negative, and sum to $1$ for $\mathbb{E}$ to be a valid transfer matrix.
Indeed, the commutation relations of Pauli matrices directly ensures that $(V_\alpha \otimes \bar V_\alpha )  \mathbb E (V_\alpha^\dagger \otimes V_\alpha^T) = \mathbb E$ for any $V_\alpha \in \{\mathds{1},X,Y,Z\}$. Up to physical isometry, this uniquely identifies the following MPS tensor:
\begin{equation} \label{eq:Achi2}
    \begin{tikzpicture} [scale = 1, baseline = {([yshift=-.5ex]current bounding box.center)}] 
    \draw[color = black, thick] (-0.8, 0) -- (0.8, 0);
    \draw[color = black, thick] (0, 0) -- (0, -0.8);
    \draw[fill = lightdodgerblue] (0,0) circle (0.3);
    \node at (-0.0,0) {\small $A$};
    \node at (0.2, -0.75) {\small $i$};
    \end{tikzpicture} = \sqrt{\lambda_i}\ \  \begin{tikzpicture} [scale = 1, baseline = {([yshift=-.5ex]current bounding box.center)}]
        \draw[color = black, thick] (-0.8, 0) -- (0.8, 0);
        \draw[fill = white] (0,0) circle (0.3);
        \node at (0,0) {\small $\sigma^i$};
        \node at (0.0, -0.79) {\small $\ $};
    \end{tikzpicture}
\end{equation}
where $\sigma^i \in \{\mathds{1}, X, Y, Z\}$ labels the Pauli matrices\footnote{In our companion work \cite{sahay2024framework}, we show that this captures all preparable $\chi=2$ MPS where $V_\alpha$ pushes through uniformly.}.

\begin{figure}
    \centering
    \includegraphics[width = 247 pt]{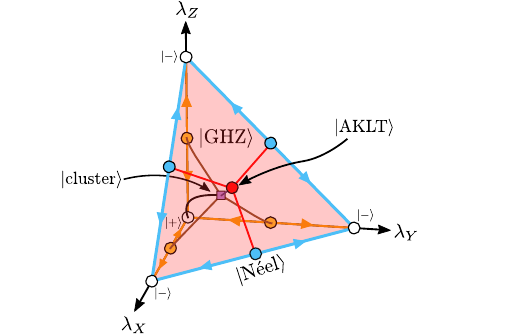}\caption{\textbf{Preparable Phase Diagram.} The solid region inside the (non-regular) tetrahedron labels a family of $\chi=2$ matrix product states which can be deterministically prepared using a single round of measurements \eqref{eq:Achi2}. This provides a rich landscape containing not only the AKLT, cluster, GHZ and N\'eel states, but also various wavefunction deformations of these states which allows one to interpolate between these celebrated states. For instance, the red line corresponds to $e^{\beta \sum_n (S^z_n)^2} \ket{\textrm{AKLT}}$ whose correlation diverges with $\beta$ as it approaches the N\'eel state.}
    \label{fig:preparable_phase_diagram}
\end{figure}

Given this class of preparable matrix product states, we can readily explore its phenomenology.
Since the parameters $\lambda_i$ of the MPS are non-negative and sum to $1$, the landscape of preparable matrix product states corresponds to points in the interior and boundary of the (non-regular) tetrahedron shown in Fig.~\ref{fig:preparable_phase_diagram}.
Furthermore, since $\sigma^{x, y, z}$ in Eq.~\eqref{eq:Achi2} can be permuted amongst themselves via a unitary MPS gauge transformation, we can always choose $\lambda_{x}\geq \lambda_y \geq \lambda_z$ (though we show the full phase diagram for ease of visualization) .

Our guided tour of this landscape begins with by noting that the celebrated AKLT chain \cite{AKLT_original} corresponds to the point $\mathbf{\lambda} = \left(0, \frac{1}{3}, \frac{1}{3}, \frac{1}{3} \right)$ on the tetrahedron, where we imagine embedding the spin-1 into a four-state qudit. While its measurement-based preparation was first discussed in Ref.~\onlinecite{smith2024constant}, we point out that there are a number of interesting deformations of the AKLT chain that can be prepared.
First, $e^{\beta \sum_n (S^z_n)^2} \ket{\text{AKLT}}$ systematically suppresses the $S^z = 0$ component of the state, which decreases $\lambda_z$ in Eq.~\eqref{eq:Achi2} (red lines in Fig.~\ref{fig:preparable_phase_diagram}).
This increases the correlation length of the AKLT chain whilst remaining in a non-trivial symmetry-protected topological (SPT) phase \cite{Gu09,Schuch2011,Pollmann_2010,Turner2011-zi,fidkowski2011topological,Chen2011-et,son12,Chen2011-kz,Pollmann2012-lv,Chen2012-oa} and in fact interpolates it to the long-range entangled N\'eel antiferromagnetic cat state $\ket{\text{N\'eel}} = \ket{\uparrow \downarrow \cdots} + \ket{\downarrow \uparrow \cdots}$ (written in the basis of $S^z = \pm 1$ states $\ket{\uparrow}, \ket{\downarrow}$ and shown as the blue points in Fig.~\ref{fig:preparable_phase_diagram}).
More generally, the AKLT chain has the remarkable property that the deformed wavefunction $\prod_{n} M_n \ket{\text{AKLT}}$ for an \emph{arbitrary} invertible complex matrix $M$ can still be deterministically prepared \cite{SM}!
This corresponds to the face of the tetrahedron, shaded in red.

Moving inward from the AKLT chain, we note that the famed cluster state \cite{Briegel01,son12,geraedts2014exact,Santos15} is located at $\mathbf{\lambda} = \left(\frac{1}{4}, \frac{1}{4}, \frac{1}{4}, \frac{1}{4} \right)$ (purple point in Fig.~\ref{fig:preparable_phase_diagram}).
Not only does the tetrahedron provide a ``measurement preparable'' path between the AKLT and cluster state, once one arrives at the cluster state, there are a number of physically motivated paths to interesting quantum states.
For example, to go the trivial $\ket{+}$ state (white dot at the origin), one travels along the $e^{\beta \sum_{n} X_n} \ket{\text{cluster}}$ (black line). 
Moreover, to go to the long-range entangled GHZ state (orange dot on the axes), one can instead travel along $e^{\beta \sum_{n} X_{2n}} \ket{\text{cluster}}$ (brown lines). 
The axes themselves correspond to deformed GHZ states, $e^{\pm \beta \sum_n X_n} \ket{\textrm{GHZ}}$ (orange lines).
Lastly, we remark that the points where the correlation length diverges can be interpreted as quantum critical points into new phases of matter. 
For instance, $e^{\beta \sum_{n} X_{2n} }\ket{\text{cluster}}$ can be transformed into $e^{\beta \sum_{n} Z_{2n-1} Z_{2n+1}} \ket{+}^{\otimes N}$ by a finite-depth local unitary, both of which limit to the same GHZ state which can be interpreted as a critical point between the trivial and cluster SPT phase \cite{Wolf06,skeleton}.

More generally, one can choose the coefficients of Eq.~\eqref{eq:Achi2} in order to engineer desired physical properties. We will explain this more constructively in the set-up for arbitary bond dimension $\chi$. For now we note that one rigid property of Eq.~\eqref{eq:Achi2} is that the bipartition entanglement spectrum of Eq.~\eqref{eq:Achi2} is always $\Lambda^2 = \left(\frac{1}{2}, \frac{1}{2}\right)$. This degeneracy is related to the state's ability to move arbitrary Pauli matrices through its virtual space. If one considers cases where $V_\alpha$ pushes through in a \emph{non-uniform} way,
the entanglement spectrum need not be flat \cite{sahay2024framework}.

\textbf{Arbitrary bond dimension.} The above construction can be generalized to any unitary error basis with the group-like property $V_g V_h  =\omega(g,h) V_{gh}$. These arise from projective representations of finite groups $G$ (with cocycle $\omega(g,h) \in U(1)$) and are also called \emph{nice error bases} \cite{knill1996group}. This gives a rich landscape even including non-abelian groups. While we give the general formula in the Supplemental Materials, here we state it for the $\chi$-dimensional clock matrices $\{ \mathcal X^a \mathcal Z^b \}$ which generalize the Pauli matrices, constituting a projective representation of $G = \mathbb Z_\chi \times \mathbb Z_\chi$.

We generalize Eq.~\eqref{eq:Achi2} to the following MPS tensor:
\begin{equation} \label{eq:AchiN}
    \begin{tikzpicture} [scale = 1, baseline = {([yshift=-.5ex]current bounding box.center)}] 
    \draw[color = black, thick] (-0.8, 0) -- (0.8, 0);
    \draw[color = black, thick] (0, 0) -- (0, -0.8);
    \draw[fill = lightdodgerblue] (0,0) circle (0.3);
    \node at (-0.0,0) {\small $A$};
    \node at (0.3, -0.75) {\small $a,b$};
    \end{tikzpicture} = \sqrt{\lambda_{a,b}}\ \  \begin{tikzpicture} [scale = 1, baseline = {([yshift=-.5ex]current bounding box.center)}]
        \draw[color = black, thick] (-0.8, 0) -- (0.8+0.8, 0);
        \draw[fill = white] (0,0) circle (0.3);
        \node at (0,0) {\small $\mathcal X^a$};
        \draw[fill = white] (0.8,0) circle (0.3);
        \node at (0.8,0) {\small $\mathcal Z^b$};
        \node at (0.0, -0.79) {\small $\ $};
    \end{tikzpicture}
\end{equation}
with $a,b \in \{1,2,\cdots,\chi\}$ labeling a $\chi^2$-dimensional qudit. Up to physical change of basis, we can presume $\lambda_{a,b} \geq 0$; moreover normalization dictates $\sum_{a,b} \lambda_{a,b}=1$. Similar to before, this state is deterministically preparable in constant-depth by measuring in the generalized Bell basis defined by the clock matrices.

\textbf{Build-your-own.} We stress that in Eq.~\eqref{eq:AchiN} one can choose $\lambda_{a,b}$ freely in order to obtain a preparable MPS with bond dimension $\chi$. In particular, if one wishes to specify eigenvalues $\mu_{a,b}$ of the transfer matrix (which dictate the correlation lengths $\xi_{a,b} = |\ln|\mu_{a,b}||^{-1}$), this can be done by setting\footnote{This follows from noting that the clock matrices are the eigenvectors of the transfer matrix.}:
\begin{equation}
\lambda_{c,d} = \frac{1}{\chi^2} \sum_{a,b} \mu_{a,b} \; \omega^{ad-bc},
\end{equation}
where $\omega=e^{2\pi i/\chi}$ and one has the condition $\lambda_{c,d} \geq 0$. Moreover, the reduced density matrix of a single $\chi^2$-state qudit has $\{ \lambda_{a,b} \}_{a,b=1,\cdots,\chi}$ as its entanglements spectrum. This follows from Eq.~\eqref{eq:AchiN} being in canonical form with a flat bipartition entanglement spectrum, such that:
\begin{equation}
    \rho = \cdots \begin{tikzpicture}[scale = 0.8, baseline = {([yshift=-.5ex]current bounding box.center)}] 
    \draw[color = black] (-0.5,0) -- (4.5, 0);
    \draw[color = black] (-0.5,-1) -- (4.5, -1);
    \foreach \i in {0, 1, 3, 4}{
        \draw[color = black] (1*\i,0) -- (1*\i, -0.5);
        \draw[fill = lightdodgerblue] (1*\i,0) circle (0.25);
        \node at (1*\i, 0) {\small $A$};
        \draw[color = black] (1*\i,-1) -- (1*\i, -0.5);
        \draw[fill = lightdodgerblue] (1*\i,-1) circle (0.25);
        \node at (1*\i, -1) {\small $\bar A$};
    }
    \draw[color = black] (1*2,0) -- (1*2, 0.5);
    \draw[fill = lightdodgerblue] (1*2,0) circle (0.25);
    \node at (1*2, 0) {\small $A$};
    \draw[color = black] (1*2,-1) -- (1*2, -1.5);
    \draw[fill = lightdodgerblue] (1*2,-1) circle (0.25);
    \node at (1*2, -1) {\small $\bar A$};
\end{tikzpicture} \cdots  =  \frac{1}{\chi}\ \
    \begin{tikzpicture}[scale = 0.8, baseline = {([yshift=-.5ex]current bounding box.center)}] 
    \draw[color = black] (2.5, 0) -- (1.5, 0);
    \draw[color = black] (2.5, -1) -- (1.5, -1);
    \draw[color = black] (2.5, -1) -- (2.5, 0);
    \draw[color = black] (1.5, -1) -- (1.5, 0);
\draw[color = black] (1*2,0) -- (1*2, 0.5);
    \draw[fill = lightdodgerblue] (1*2,0) circle (0.25);
    \node at (1*2, 0) {\small $A$};
    \draw[color = black] (1*2,-1) -- (1*2, -1.5);
    \draw[fill = lightdodgerblue] (1*2,-1) circle (0.25);
    \node at (1*2, -1) {\small $\bar A$};
    \end{tikzpicture}
\end{equation}
We thus see that efficiently preparable states can have very rich correlation functions and entanglement properties.

\textbf{Higher-dimensional examples.} To illustrate how these ideas can be used to prepare states in higher dimensions, let us consider a state $\ket{\Psi(\beta)}$ which interpolates from a trivial product state ($\beta=0$) to the square lattice toric code \cite{Kitaev_2003} ($\beta \to \infty$). In particular, $\ket{\Psi(\beta)} \propto e^{\beta \sum_v A_v} \ket{+}^{\otimes N}$ where $A_v = \prod_{l \in v} Z_l$ and $B_p = \prod_{l \in p} X_l$ are the `star' and `plaquette' stabilizers of the toric code. We can prepare this state using measurement by starting with the following $\chi=2$ PEPS tensor:
\begin{equation}
\includegraphics[valign = c, scale = 0.6]{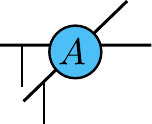} \; = \; \includegraphics[valign = c, scale = 0.6]{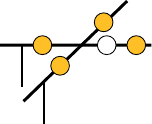}
\end{equation}
where the orange circles are Hadamard matrices, the white circle is $e^{\alpha Z}$ where $\tanh \alpha = e^{-2\beta}$, and the `$T$' and `$X$'-junctions are to be understood as Kronecker deltas $\delta_{ijk}$ and $\delta_{ijkl}$ respectively. We can follow our general prescription by noticing that the Pauli's again form a unitary error basis with nice push-through rules:
\begin{equation}
\includegraphics[valign = c, scale = 0.55]{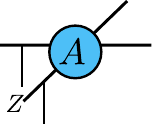}
= \includegraphics[valign = c, scale = 0.55]{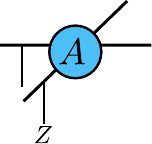}, \quad 
\includegraphics[valign = c, scale = 0.55]{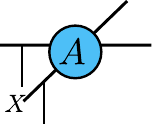}
= \includegraphics[valign = c, scale = 0.55]{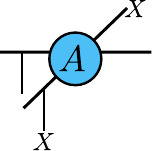},
\end{equation}
where $X$ can take any turn. This way, the $Z$ errors can be locally corrected, whereas the $X$ errors can be pushed through to infinity, or more efficiently pairwise annihilated wherever they are found. We note that upon acting with a finite-depth local unitary circuit, this measurement-preparable path constitutes a phase transition between a trivial and higher-form SPT phase, where the toric code arises as a gapless critical point \cite{pivot}.

Our second example is $\ket{\Psi(\beta)} \propto e^{\beta \sum_v X_v} |\textrm{GHZ}\rangle$ on the square lattice. In principle we could prepare this via a two-step process by applying the gauging map \cite{LREfromSPT} to the previous example. However, it is interesting to see that it is \emph{directly} preparable. Moreover, the push-through rules are quite different, potentially suggesting different generalizations. The local tensor for $\beta=0$ is simply the Kronecker delta, and $A(\beta)$ is then given by appending $e^{\beta X}$ to the physical leg. We read off the following push-through rules:
\begin{equation}
\includegraphics[valign = c, scale = 0.55]{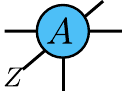}
= \includegraphics[valign = c, scale = 0.55]{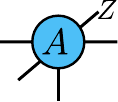},\quad
\includegraphics[valign = c, scale = 0.55]{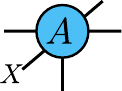}
= \includegraphics[valign = c, scale = 0.55]{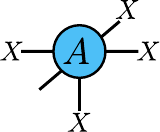},
\end{equation}
where $Z$ can again take any turn on the virtual bonds. One might worry about a single $X$ spawning into three other virtual $X$ insertions. This means that an \emph{odd} number of $X$'s on a single plaquette cannot be removed. Fortunately, this has zero probability of happening, since the corresponding tensor network contracts to zero. Since we are guaranteed an even number of $X$ insertions per plaquette, they form closed loops such that we can correct them by acting on the enclosed domains.

\textbf{Diagnostics.} The fact that the landscape of preparable tensor network states is so rich, naturally raises the question: how do we test whether a given tensor network can be prepared in this way? The key condition is Eq.\eqref{eq:VTV=T}, which can be interpreted as a generalized eigenvalue equation. It is most instructive to interpret $\mathbb E$ as a matrix in the \emph{vertical} direction, where we think of the top virtual bonds as input and the bottom bonds as output. Then $\mathbb E$ is a positive semi-definite map and can be diagonalized as $\mathbb U^\dagger \mathbb D \mathbb U$ where $\mathbb D \geq 0$. Hence, if $\mathbb V_\alpha$ denotes the tensor product of $V_\alpha$ operators\footnote{I.e., $\mathbb V_\alpha = V_\alpha \otimes (V_\alpha^l)^\dagger \otimes (V_\alpha^r)^\dagger \otimes (V_\alpha^f)^\dagger$.} in Eq.~\eqref{eq:VTV=T}, then the space of solutions is spanned by $\mathbb V_\alpha = \mathbb U^\dagger \mathbb D_\alpha^\dagger \mathbb U$ where $\{\mathbb D_\alpha\}$ are matrices which commute with $\mathbb D$. Thus in the generic case where $\mathbb D$ is a non-degenerate diagonal matrix, then $\{ \mathbb D_\alpha \}$ is simply given by the $\chi^z$ independent choices of diagonal matrices (where $z$ is the number of virtual bonds; $z=2$ for MPS).

The crux of the matter is to then find \emph{disentangled} states within this vector space of solutions; indeed those correspond to tensor product operators on the separate virtual legs in Eq.~\eqref{eq:VTV=T}. In the most general case, this can be achieved by a non-linear optimization within this space, which we have bench-marked for several of our examples. Alternatively, one can employ a bootstrap-inspired approach: if Eq.~\eqref{eq:VTV=T} admits disentangled solutions, this implies a non-trivial constraint on the set of solutions after blocking several sites. Hence, disentangled solutions can be found by obtaining the intersection of these solution spaces.

In a subset of cases, such as the one-dimensional examples we studied where $V_\alpha$ pushes through uniformly and commute up to a phase, the unitary error basis is given by the columns of $\mathbb U$ itself. This is easiest to see by noting that Eq.~\eqref{eq:Achi2} and Eq.~\eqref{eq:AchiN} are already in this diagonalized form. Indeed, $\mathbb U$ being a unitary super-operator is equivalent to its columns defining an error basis. Hence, this shows that in certain cases, one can claim success after a finite number of steps. Moreover, in certain cases one can conclude a state is \emph{not} preparable: our companion work \cite{sahay2024framework} identifies a no-go theorem in a one-dimensional setting based on properties of the entanglement spectrum and correlation functions.

\textbf{Generalizations.} Thus far, we have discussed the preparation of interesting families of one- and two-dimensional tensor networks by `fusing' or `gluing' their virtual legs together by one-round of complete basis measurement.
Here, we provide first steps for generalizations where we permit \emph{multiple} rounds of measurements and feedback, as well as exploring the preparation and application of tensor network \emph{operators}.
To do so, let us recognize that, since the cluster state corresponds to point $\mathbf{\lambda} = \left(\frac{1}{4}, \frac{1}{4}, \frac{1}{4}, \frac{1}{4} \right)$ of Eq.~\eqref{eq:Achi2}, it is preparable from one-round of measurements and tensor product unitary feedback\footnote{We remark in our companion work \cite{sahay2024framework} that, it can in fact be prepared using a one-site unit cell by leveraging its properties as a dipole SPT \cite{han2024dipole, lam2024classification}.}.
However, we may re-interpret the cluster state's MPS representation as a matrix product operator (MPO) via: 
\begin{equation} \label{eq-clusterMPO}
    \begin{tikzpicture} [scale = 1.5, baseline = {([yshift=-.5ex]current bounding box.center)}] 
        \draw[color = black, thick] (-1.25, 0) -- (0.75, 0);
        \draw[color = black, thick] (-1, 0) -- (-1, -0.3);
        \draw[color = black, thick] (-0.5, 0) -- (-0.5, -0.3);
        \draw[color = black, thick] (0, 0) -- (0, -0.3);
        \draw[color = black, thick] (0.5, 0) -- (0.5, -0.3);
        \draw[fill = lightdodgerblue, thick] (-1, 0) circle (0.1);
        \draw[fill = lightdodgerblue, thick] (-0.5, 0) circle (0.1);
        \draw[fill = lightdodgerblue, thick] (-0, 0) circle (0.1);
        \draw[fill = lightdodgerblue, thick] (0.5, 0) circle (0.1);
    \end{tikzpicture} \longrightarrow     \begin{tikzpicture} [scale = 1.5, baseline = {([yshift=-.5ex]current bounding box.center)}] 
        \draw[color = black, thick] (-1.25, 0) -- (0.75, 0);
        \draw[color = black, thick] (-1, 0) -- (-1, -0.3);
        \draw[color = black, thick] (-0.5, 0) -- (-0.5, 0.3);
        \draw[color = black, thick] (0, 0) -- (0, -0.3);
        \draw[color = black, thick] (0.5, 0) -- (0.5, 0.3);
        \draw[fill = lightdodgerblue, thick] (-1, 0) circle (0.1);
        \draw[fill = lightdodgerblue, thick] (-0.5, 0) circle (0.1);
        \draw[fill = lightdodgerblue, thick] (-0, 0) circle (0.1);
        \draw[fill = lightdodgerblue, thick] (0.5, 0) circle (0.1);
    \end{tikzpicture}
\end{equation}
In recent years, this MPO has been appreciated as representing the Kramers-Wannier ``gauging'' map \cite{LREfromSPT, hierarchy, seiberg2024noninvertible, Aasen_2016, lootens_dualities_2023, lootens_dualities_2024, haegeman_2015_gauging}. 
Ref.~\onlinecite{LREfromSPT} showed how the action of this MPO can be deterministically implemented on a given input state using a finite-depth measurement-based protocol.
Here we offer a re-interpretation, viewing the cluster state in Eq.~\eqref{eq-clusterMPO} as a ``stored-program'' \cite{Gottesman_1999, Raussendorf2001-xm, Briegel_2009}, which is prepared first and then applied to a given many-body state using measurements.
In particular, given an initial many-body state (say, prepared with one-round of measurements and feedback), we can apply the above MPO (blue) to the state (orange) as:
\begin{equation}
    \begin{tikzpicture} [scale = 1.5, baseline = {([yshift=-.5ex]current bounding box.center)}] 
        \draw[color = black, thick] (-1.25, 0) -- (0.75, 0);
        \draw[color = black, thick] (-1, 0) -- (-1, -0.3);
        \draw[color = black, thick] (-0.5, 0) -- (-0.5, 0.3);
        \draw[color = black, thick] (0, 0) -- (0, -0.3);
        \draw[color = black, thick] (0.5, 0) -- (0.5, 0.3);
        \draw[fill = lightdodgerblue, thick] (-1, 0) circle (0.1);
        \draw[fill = lightdodgerblue, thick] (-0.5, 0) circle (0.1);
        \draw[fill = lightdodgerblue, thick] (-0, 0) circle (0.1);
        \draw[fill = lightdodgerblue, thick] (0.5, 0) circle (0.1);
        \draw[color = black, thick] (-1.25, 0 + 0.8) -- (0.75, 0 + 0.8);
        \draw[color = black, thick] (-0.5, 0 + 0.8) -- (-0.5, -0.3 + 0.8);
        \draw[color = black, thick] (0.5, 0 + 0.8) -- (0.5, -0.3 + 0.8);
        \draw[fill = lightorange(ryb), thick] (-0.5, 0.8) circle (0.1);
        \draw[fill = lightorange(ryb), thick] (0.5, 0.8) circle (0.1);
        \draw[fill = lightishgray, thick] (0.5, 0.4) circle (0.1);
        \draw[fill = lightishgray, thick] (-0.5, 0.4) circle (0.1);
    \end{tikzpicture}
\end{equation}
where the gray circles indicate the measurements performed between one-sublattice of the cluster state and the qubits of the original state.
One can readily show that undesired measurement outcomes can always be corrected at the physical level in such a case \cite{LREfromSPT}.
This can be viewed as a generalization of the standard teleportation protocol \cite{Teleportation}, where the resource is a cluster state instead of decoupled Bell pairs.

Lastly, another natural extension is to only use \emph{incomplete} basis measurements. This can be naturally motivated in cases where some of the error correction is local. E.g., in the toric code example we saw that $Z$ errors can be locally corrected. This means that the feedback protocol can be turned into a logical quantum gate, obviating the need for the corresponding measurement. In other words, the toric code example can be prepared\footnote{This observation was also made in Ref.~\onlinecite{lu2022shortcut} for deterministically preparing the fixed-point toric code.} using only a `$ZZ= \pm 1$' measurement between the two ancillas (rather than a full Bell basis measurement which includes $XX = \pm 1$) \cite{SM}. In our companion work we also highlight how even in the one-dimensional case this can be utilized to prepare states whose entanglement spectrum is not flat \cite{sahay2024framework}.

In conclusion, the submanifold of tensor network states which can be efficiently prepared using measurement invites further exploration. It exhibits remarkably rich phenomenology, whilst at the same time displaying tight connections between its entanglement spectra and correlation functions---which need not hold for more generic tensor network states. We have highlighted several exciting open questions, and many more remain, including complete classifications beyond those in the companion work \cite{sahay2024framework}, improved numerical algorithms for detecting preparability, charting out the physical properties (in)compatible with measurement-based preparation, and finally bringing these ideas to experimental fruition in novel platforms with mid-circuit measurement capabilities \cite{Singh_2023,moses2023race,deist2022midcircuit,norcia2023midcirc,graham2023midcirc,baumer2023efficient, anand2024dualspecies, baumer2024quantum, finkelstein2024universal, Bluvstein_2023}.

\vspace{5pt}

\emph{Note added:} The posting of this preprint to the arXiv was coordinated with simultaneous postings by Smith et al.~\cite{smith2024constant} and Stephen et al.~\cite{stephen2024fusing}. Both discuss the measurement-based preparation of matrix product states, and were developed independently from this work.

\section{Acknowledgements} We would like to thank Richard Allen, Soonwon Choi, Kevin Smith, and Francisco Machado.
R.S. acknowledges support from the U.S. Department of Energy, Office of Science, Office of Advanced Scientific Computing Research, Department of Energy Computational Science Graduate Fellowship under Award Number
DESC0022158.
R.V. is supported by the
Simons Collaboration on Ultra-Quantum Matter, which
is a grant from the Simons Foundation (618615, Ashvin
Vishwanath).

\bibliography{refs}

\pagebreak
\onecolumngrid
\appendix

\begin{center}
{\large \textbf{Supplementary Materials}}    
\end{center}

\section{Supplementary Materials A: Constructive Proof for Physical Unitary}

\begin{shaded}
    \noindent
    \textbf{Theorem} (Virtual-Physical Correspondence) Suppose $A$ is a tensor with physical Hilbert space dimension $d$ and every virtual leg has dimension $\chi$.
    Below, we show this tensor as a matrix product state but we remark that the theorem statement and proof are agnostic to dimension.
    Now, suppose that $V$ is an operator acting on the virtual legs.
    Then, there exists a unitary $U$ such that:
    \begin{equation}\label{eq-UtoV}
    \begin{tikzpicture} [scale = 1, baseline = {([yshift=-.5ex]current bounding box.center)}] 
    \draw[color = black] (-0.8, 0) -- (0.8, 0); 
    \draw[color = black] (0, 0) -- (0, -0.8); 
    \draw[color = black] (-0.8, 0) -- (-0.8, 1);
    \draw[color = black] (0.8, 0) -- (0.8, 1);
    \draw[rounded corners, fill = lightorange(ryb)] (-1, 0.35) rectangle (1, 0.85);
    \draw[fill = lightdodgerblue] (0,0) circle (0.3); 
    \node at (-0.03,0) {\small $A$};
    \node at (0, 0.58) {\small $V$};
    \end{tikzpicture}
     =     \begin{tikzpicture} [scale = 1, baseline = {([yshift=-.5ex]current bounding box.center)}] 
    \draw[color = black] (-0.8, 0) -- (0.8, 0); 
    \draw[color = black] (0, 0) -- (0, -0.8); 
    \draw[fill = lightdodgerblue] (0,0) circle (0.3); 
    \node at (-0.03,0) {\small $A$};
    \node at (0.0, -1) {\small $U$};
    \end{tikzpicture}
\end{equation}
    if and only if, $V$ leaves the transfer operator invariant:
    \begin{equation}
    \begin{tikzpicture} [scale = 1, baseline = {([yshift=-.5ex]current bounding box.center)}] 
    \draw[color = black] (-0.8, 0) -- (0.8, 0); 
    \draw[color = black] (-0.8, 0) -- (-0.8, 1);
    \draw[color = black] (0.8, 0) -- (0.8, 1);
    \draw[color = black] (-0.8, -1.2) -- (-0.8, -2.2);
    \draw[color = black] (0.8, -1.2) -- (0.8, -2.2);
    \draw[color = black] (-0.8, -1.2) -- (0.8, -1.2);
    \draw[rounded corners, fill = lightorange(ryb)] (-1, 0.35) rectangle (1, 0.85);
    \draw[rounded corners, fill = lightorange(ryb)] (-1, -1.2 - 0.35) rectangle (1, -1.2-0.85);
    \draw[color = black] (0, 0) -- (0, -1.2); 
    \draw[fill = lightdodgerblue] (0,0) circle (0.3); 
    \draw[fill = lightdodgerblue] (0,-1.2) circle (0.3); 
    \node at (-0.03,0) {\small $A$};
    \node at (-0.03,-1.2) {\small $\bar{A}$};
    \node at (0, 0.58) {\small $V$};
    \node at (0, -1.2 - 0.58) {\small $V^{\dagger}$};
    %
    \end{tikzpicture} = \begin{tikzpicture} [scale = 1, baseline = {([yshift=-.5ex]current bounding box.center)}] 
    \draw[color = black] (-0.8, 0) -- (0.8, 0); 
    \draw[color = black] (-0.8, -1.2) -- (0.8, -1.2); 
    \draw[color = black] (0, 0) -- (0, -1.2); 
    \draw[fill = lightdodgerblue] (0,0) circle (0.3); 
    \draw[fill = lightdodgerblue] (0,-1.2) circle (0.3); 
    \node at (-0.03,0) {\small $A$};
    \node at (-0.03,-1.2) {\small $\bar{A}$};
    %
    \end{tikzpicture}
    \end{equation}
    Moreover, $U$ can take the form: 
    \begin{equation}
    U\,  =\,   
    \begin{tikzpicture} [scale = 1, baseline = {([yshift=-.5ex]current bounding box.center)}] 
    \draw[color = black] (-0.8, 0) -- (0.8, 0); 
    \draw[color = black] (0, 0) -- (0, 0.5); 
    \draw[fill = lightdodgerblue] (0,0) circle (0.24); 
    \node at (0.0,0) {\small $A^+$};
    %
    %
    \draw[color = black] (-0.8, 0) -- (-0.8, -1);
    \draw[color = black] (0.8, 0) -- (0.8, -1);
    \draw[color = black] (-0.8, -1) -- (0.8, -1); 
    \draw[color = black] (0, -1) -- (0, -0.5-1); 
    \draw[fill = lightdodgerblue] (0,-1) circle (0.24); 
    \draw[rounded corners, fill = lightorange(ryb)] (-1, -0.7) rectangle (1, -0.3);
    \node at (0.,-1) {\small $A$};
    \node at (0,-0.5) {\small $V$};
    \end{tikzpicture} + \left(\mathds{1}_{d}  - \begin{tikzpicture} [scale = 1, baseline = {([yshift=-.5ex]current bounding box.center)}] 
    \draw[color = black] (-0.8, 0) -- (0.8, 0); 
    \draw[color = black] (0, 0) -- (0, 0.5); 
    \draw[fill = lightdodgerblue] (0,0) circle (0.24); 
    \node at (0.0,0) {\small $A^+$};
    %
    %
    \draw[color = black] (-0.8, 0) -- (-0.8, -1);
    \draw[color = black] (0.8, 0) -- (0.8, -1);
    \draw[color = black] (-0.8, -1) -- (0.8, -1); 
    \draw[color = black] (0, -1) -- (0, -0.5-1); 
    \draw[fill = lightdodgerblue] (0,-1) circle (0.24); 
    \node at (0.,-1) {\small $A$};
    \end{tikzpicture} \right)
    \end{equation}
where $\hat{A}$ is the Moore-Penrose pseudo-inverse of $A$ (viewed as a map from virtual to physical bonds). 
%
\end{shaded}

\textit{Proof.} Let us remark that the forwards direction of this proof is trivial.
Therefore, we focus our attention on the backwards direction.
We prove this by explicit construction of $U$.
To do so, it is convenient and illuminating to perform a singular value decomposition on the matrix product state as:
\begin{equation}
         \begin{tikzpicture} [scale = 1, baseline = {([yshift=-.5ex]current bounding box.center)}] 
    \draw[color = black] (-0.8, 0) -- (0.8, 0); 
    \draw[color = black] (0, 0) -- (0, -0.8); 
    \draw[fill = lightdodgerblue] (0,0) circle (0.3); 
     \node at (-0.03,0) {\small $A$};
    \end{tikzpicture} =          \begin{tikzpicture} [scale = 1, baseline = {([yshift=-.5ex]current bounding box.center)}] 
    \draw[color = black] (-0.8, 0) -- (0.8, 0); 
    \draw[color = black] (0, 0) -- (0, -1.4); 
    \draw[fill = lightdodgerblue] (0,0) circle (0.24); 
    \draw[fill = white] (0,-0.5) circle (0.14);
    \draw[fill = lightorange(ryb)] (0,-1) circle (0.24);
    \node at (0.02,0) {\small $\mathcal{U}$};
    \node at (0.0,-0.5) {\small $s$};
    \node at (0.0,-1) {\small $\mathcal{V}^{\dagger}$};
    \end{tikzpicture} \to           \begin{tikzpicture} [scale = 1, baseline = {([yshift=-.5ex]current bounding box.center)}] 
    \draw[color = black] (-0.8, 0) -- (0.8, 0); 
    \draw[color = black] (0, 0) -- (0, -0.8); 
    \draw[fill = lightdodgerblue] (0,0) circle (0.24); 
    \draw[fill = white] (0,-0.5) circle (0.14);
    \node at (0.02,0) {\small $\mathcal{U}$};
    \node at (0.0,-0.5) {\small $s$};
    \end{tikzpicture}
\end{equation}
where in the last step we dropped the $V^{\dagger}$ because it can always be removed from the matrix product state by a physical local unitary transformation.
We will call the above the ``vertical representation'' of the MPS tensor (for lack of a better name and because I'll refer to it from time to time).
Vertical representation in hand, we consider the following trial unitary operator: 
\begin{equation}
    U\,  =\,   \begin{tikzpicture} [scale = 1, baseline = {([yshift=-.5ex]current bounding box.center)}] 
    \draw[color = black] (-0.8, 0) -- (0.8, 0); 
    \draw[color = black] (0, 0) -- (0, 0.8); 
    \draw[fill = lightdodgerblue] (0,0) circle (0.24); 
    \draw[fill = white] (0,0.5) circle (0.14);
    \node at (0.02,0) {\small $\bar{\mathcal{U}}$};
    \node at (0.1 ,0.55) {\small $s^+$};
    \draw[color = black] (-0.8, 0) -- (-0.8, -1);
    \draw[color = black] (0.8, 0) -- (0.8, -1);
    \draw[color = black] (-0.8, -1) -- (0.8, -1); 
    \draw[color = black] (0, -1) -- (0, -0.8-1); 
    \draw[fill = lightdodgerblue] (0,-1) circle (0.24); 
    \draw[fill = white] (0,-0.5-1) circle (0.14);
    \draw[rounded corners, fill = lightorange(ryb)] (-1, -0.7) rectangle (1, -0.3);
    \node at (0.02,-1) {\small ${\mathcal{U}}$};
    \node at (0,-0.5) {\small $V$};
    \node at (0.0 ,-0.5-1) {\small $s$};
    \end{tikzpicture} + \left(1 -\ \begin{tikzpicture} [scale = 1, baseline = {([yshift=-.5ex]current bounding box.center)}] 
    \draw[color = black] (0.0, 0.0) -- (0.0, -1);
    \draw[fill = white] (0,-0.75) circle (0.14);
    \draw[fill = white] (0,-0.25) circle (0.14);
    \node at (0.0 ,-0.75) {\small $s$};
    \node at (0.1 ,-0.175) {\small $s^+$};
    \end{tikzpicture} \right) = X + (1 - P)
\end{equation}
where we introduced the notation $s^+$ to refer to the Moore-Penrose pseudo-inverse of the singular value matrix.
%
%
Operationally, the Moore-Penrose pseudo-inverse (for diagonal matrices) is obtained by inverting each of the non-zero elements along the diagonal of $s$ and then taking the transpose.
It is worth noting some special properties of $s^+$:
\begin{enumerate}
    \item[1.] $s^+$ and $s$ are weak inverses of one another.
    That is to say that $s^+ s s^+ = s^+$ and similarly $ss^+ s = s$.

    \item[2.] $s^+ s$ and $s s^+$ are both projectors (i.e. are idempotent and hermitian).
    In the case, where $s$ is injective (i.e. full rank), $s^+ s = 1$.    
\end{enumerate}
Note that the second property guarantees that $(1 - P)$ is a projector.
With these properties, we can prove first that $U$ is unitary and can be pushed through $A$ to yield $V$.
To prove unitarity, note that:
\begin{equation}
    U^{\dagger} U = X^{\dagger} X + (1 - P) + (1 - P)X + X^{\dagger}(1 - P) 
\end{equation}
We evaluate each term above individually.
Let us first note that: 
\begin{equation}
    PX =  
    \begin{tikzpicture} [scale = 1, baseline = {([yshift=-.5ex]current bounding box.center)}] 
    \draw[color = black] (-0.8, 0) -- (0.8, 0); 
    \draw[color = black] (0, 0) -- (0, 0.7); 
    \draw[fill = lightdodgerblue] (0,0) circle (0.24); 
    \draw[fill = white] (0, 0.5) circle (0.14);
    \node at (0.02,0) {\small $\bar{\mathcal{U}}$};
    \node at (0.1 ,0.55) {\small $s^+$};
    \draw[color = black] (-0.8, 0) -- (-0.8, -1);
    \draw[color = black] (0.8, 0) -- (0.8, -1);
    \draw[color = black] (-0.8, -1) -- (0.8, -1); 
    \draw[color = black] (0, -1) -- (0, -1.5-1); 
    \draw[fill = lightdodgerblue] (0,-1) circle (0.24); 
    \draw[fill = white] (0,-0.5-1) circle (0.14);
    \draw[fill = white] (0,-1-0.85) circle (0.14);
    \draw[fill = white] (0,-1-1.2) circle (0.14);
    \draw[rounded corners, fill = lightorange(ryb)] (-1, -0.7) rectangle (1, -0.3);
    \node at (0.02,-1) {\small $\mathcal{U}$};
    \node at (0,-0.5) {\small $V$};
    \node at (0.0 ,-0.5-1) {\small $s$};
    \node at (0.1 ,-1-0.80) {\small $s^+$};
    \node at (0.0 ,-1-1.2) {\small $s$};
     \end{tikzpicture} =\begin{tikzpicture} [scale = 1, baseline = {([yshift=-.5ex]current bounding box.center)}] 
    \draw[color = black] (-0.8, 0) -- (0.8, 0); 
    \draw[color = black] (0, 0) -- (0, 0.8); 
    \draw[fill = lightdodgerblue] (0,0) circle (0.24); 
    \draw[fill = white] (0,0.5) circle (0.14);
    \node at (0.02,0) {\small $\bar{\mathcal{U}}$};
    \node at (0.1 ,0.55) {\small $s^+$};
    \draw[color = black] (-0.8, 0) -- (-0.8, -1);
    \draw[color = black] (0.8, 0) -- (0.8, -1);
    \draw[color = black] (-0.8, -1) -- (0.8, -1); 
    \draw[color = black] (0, -1) -- (0, -0.8-1); 
    \draw[fill = lightdodgerblue] (0,-1) circle (0.24); 
    \draw[fill = white] (0,-0.5-1) circle (0.14);
    \draw[rounded corners, fill = lightorange(ryb)] (-1, -0.7) rectangle (1, -0.3);
    \node at (0.02,-1) {\small ${\mathcal{U}}$};
    \node at (0,-0.5) {\small $V$};
    \node at (0.0 ,-0.5-1) {\small $s$};
    \end{tikzpicture}
    = X
\end{equation}
    As an aside, note that a similar calculation reveals that $PX = XP = X$.
    This implies that $U$ decomposes as a direct sum in the spaces with support on $P$ and $(1 - P)$, with $X$ only acting on the former space.
    Indeed, as a consequence of the above calculation, we have that:
    \begin{equation}
        U^{\dagger} U = X^{\dagger} X+ (1 - P)
    \end{equation}
    Now, let us note that:
    \begin{equation}
        X^{\dagger} X = \begin{tikzpicture} [scale = 1, baseline = {([yshift=-.5ex]current bounding box.center)}] 
    \draw[color = black] (-0.8, 0) -- (0.8, 0); 
    \draw[color = black] (0, 0) -- (0, 0.8); 
    \draw[fill = lightdodgerblue] (0,0) circle (0.24); 
    \draw[fill = white] (0,0.5) circle (0.14);
    \node at (0.02,0) {\small $\bar{\mathcal{U}}$};
    \node at (0.1 ,0.55) {\small $s^+$};
    \draw[color = black] (-0.8, 0) -- (-0.8, -1);
    \draw[color = black] (0.8, 0) -- (0.8, -1);
    \draw[color = black] (-0.8, -1) -- (0.8, -1); 
    \draw[color = black] (0, -1) -- (0, -1-1); 
    \draw[fill = lightdodgerblue] (0,-1) circle (0.24); 
    
    \draw[rounded corners, fill = lightorange(ryb)] (-1, -0.7) rectangle (1, -0.3);
    \node at (0.02,-1) {\small ${\mathcal{U}}$};
    \node at (0,-0.5) {\small $V$};
    \draw[fill = white] (0,-0.75-1) circle (0.14);
    \node at (0.07 ,-0.7-1) {\small $s^{\dagger}$};
    \draw[fill = white] (0,-0.4-1) circle (0.14);
    \node at (0.0 ,-0.4-1) {\small $s$};
    \draw[color = black] (-0.8, -2.2) -- (0.8, -2.2); 
    \draw[fill = lightdodgerblue] (0,-2.2) circle (0.24); 
    \node at (0.02,-2.2) {\small $\bar{\mathcal{U}}$};
    %
    %
    \draw[color = black] (-0.8, -2.2) -- (-0.8, -1-2.2);
    \draw[color = black] (0.8, -2.2) -- (0.8, -1-2.2);
    \draw[color = black] (-0.8, -1-2.2) -- (0.8, -1-2.2); 
    \draw[color = black] (0, -1-2.2) -- (0, -0.8-1-2.2); 
    \draw[fill = lightdodgerblue] (0,-1-2.2) circle (0.24); 
    \draw[rounded corners, fill = lightorange(ryb)] (-1, -0.7-2.2) rectangle (1, -0.3-2.2);
    \node at (0.02,-1-2.2) {\small ${\mathcal{U}}$};
    \node at (0,-0.5-2.2) {\small $V^{\dagger}$};
    \draw[fill = white] (0,-0.5-1 -2.2) circle (0.14);
    \node at (0.13 ,-0.47-3.2) {\small $(s^+)^{\dagger}$};
    \end{tikzpicture} = \begin{tikzpicture} [scale = 1, baseline = {([yshift=-.5ex]current bounding box.center)}] 
    \draw[color = black] (-0.8, 0) -- (0.8, 0); 
    \draw[color = black] (0, 0) -- (0, 0.8); 
    \draw[fill = lightdodgerblue] (0,0) circle (0.24); 
    \draw[fill = white] (0,0.5) circle (0.14);
    \node at (0.02,0) {\small $\bar{\mathcal{U}}$};
    \node at (0.1 ,0.55) {\small $s^+$};
    \draw[color = black] (-0.8, 0) -- (-0.8, -1);
    \draw[color = black] (0.8, 0) -- (0.8, -1);
    \draw[color = black] (-0.8, -1) -- (0.8, -1); 
    \draw[color = black] (0, -1) -- (0, -1-1); 
    \draw[fill = lightdodgerblue] (0,-1) circle (0.24); 
    
    \node at (0.02,-1) {\small ${\mathcal{U}}$};
    \draw[fill = white] (0,-0.75-1) circle (0.14);
    \node at (0.07 ,-0.7-1) {\small $s^{\dagger}$};
    \draw[fill = white] (0,-0.4-1) circle (0.14);
    \node at (0.0 ,-0.4-1) {\small $s$};
    \draw[color = black] (-0.8, -2.2) -- (0.8, -2.2); 
    \draw[fill = lightdodgerblue] (0,-2.2) circle (0.24); 
    \node at (0.02,-2.2) {\small $\bar{\mathcal{U}}$};
    %
    %
    \draw[color = black] (-0.8, -2.2) -- (-0.8, -1-2.2);
    \draw[color = black] (0.8, -2.2) -- (0.8, -1-2.2);
    \draw[color = black] (-0.8, -1-2.2) -- (0.8, -1-2.2); 
    \draw[color = black] (0, -1-2.2) -- (0, -0.8-1-2.2); 
    \draw[fill = lightdodgerblue] (0,-1-2.2) circle (0.24); 
    \node at (0.02,-1-2.2) {\small ${\mathcal{U}}$};
    \draw[fill = white] (0,-0.5-1 -2.2) circle (0.14);
    \node at (0.12 ,-0.47-3.2) {\small $(s^+)^{\dagger}$};
    \end{tikzpicture} = (s^+)^{\dagger} s^{\dagger} s s^+ = (ss^+)^{\dagger} (s s^+) = \begin{tikzpicture} [scale = 1, baseline = {([yshift=-.5ex]current bounding box.center)}] 
    \draw[color = black] (0.0, 0.0) -- (0.0, -1);
    \draw[fill = white] (0,-0.75) circle (0.14);
    \draw[fill = white] (0,-0.25) circle (0.14);
    \node at (0.0 ,-0.75) {\small $s$};
    \node at (0.1 ,-0.2) {\small $s^+$};
    \end{tikzpicture} = P
    \end{equation}
    where in the second step we used the fact that $V$ leaves the transfer matrix invariant.
    Therefore, we have that:
    \begin{equation}
        U^{\dagger} U = 1
    \end{equation}
    Now, note that $U$ is a square matrix and since $U^{\dagger}$ is a left inverse of $U$, it is also its right inverse.
    Thus, $U U^{\dagger} = 1$.
    Thus, $U$ is unitary.
    Having proven unitarity, we now prove that Eq.~\eqref{eq-UtoV} holds.
    To do so, we first note the following non-trivial identity
    \begin{equation}
        UU^{\dagger} = X X^{\dagger} + (1 - P) + X(1 - P) + (1 - P)X^{\dagger} = 1 \implies XX^{\dagger} = P 
    \end{equation}
    This yields the following diagrammatic identity:
    \begin{equation}
        XX^{\dagger} = \begin{tikzpicture} [scale = 1, baseline = {([yshift=-.5ex]current bounding box.center)}] 
    \draw[color = black] (-0.8, 0) -- (0.8, 0); 
    \draw[color = black] (0, 0) -- (0, 0.8); 
    \draw[fill = lightdodgerblue] (0,0) circle (0.24); 
    \draw[fill = white] (0,0.5) circle (0.14);
    \node at (0.02,0) {\small $\bar{\mathcal{U}}$};
    \node at (0.05 ,0.57) {\small $s^{\dagger}$};
    \draw[color = black] (-0.8, 0) -- (-0.8, -1);
    \draw[color = black] (0.8, 0) -- (0.8, -1);
    \draw[color = black] (-0.8, -1) -- (0.8, -1); 
    \draw[color = black] (0, -1) -- (0, -1-1); 
    \draw[fill = lightdodgerblue] (0,-1) circle (0.24); 
    
    \draw[rounded corners, fill = lightorange(ryb)] (-1, -0.7) rectangle (1, -0.3);
    \node at (0.02,-1) {\small ${{\mathcal{U}}}$};
    \node at (0,-0.5) {\small $V^{\dagger}$};
    \draw[fill = white] (0,-0.75-1) circle (0.14);
    \node at (0.1 ,-0.7-1) {\small $s^+$};
    \draw[fill = white] (0,-0.4-1) circle (0.14);
    \node at (0.12 ,-0.35-1) {\small $(s^+)^{\dagger}$};
    \draw[color = black] (-0.8, -2.2) -- (0.8, -2.2); 
    \draw[fill = lightdodgerblue] (0,-2.2) circle (0.24); 
    \node at (0.02,-2.2) {\small $\bar{\mathcal{U}}$};
    %
    %
    \draw[color = black] (-0.8, -2.2) -- (-0.8, -1-2.2);
    \draw[color = black] (0.8, -2.2) -- (0.8, -1-2.2);
    \draw[color = black] (-0.8, -1-2.2) -- (0.8, -1-2.2); 
    \draw[color = black] (0, -1-2.2) -- (0, -0.8-1-2.2); 
    \draw[fill = lightdodgerblue] (0,-1-2.2) circle (0.24); 
    \draw[rounded corners, fill = lightorange(ryb)] (-1, -0.7-2.2) rectangle (1, -0.3-2.2);
    \node at (0.02,-1-2.2) {\small $\mathcal{U}$};
    \node at (0,-0.5-2.2) {\small $V$};
    \draw[fill = white] (0,-0.5-1 -2.2) circle (0.14);
    \node at (0 ,-0.5-3.2) {\small $s$};
    \end{tikzpicture} = \begin{tikzpicture} [scale = 1, baseline = {([yshift=-.5ex]current bounding box.center)}] 
    \draw[color = black] (0.0, 0.0) -- (0.0, -1);
    \draw[fill = white] (0,-0.75) circle (0.14);
    \draw[fill = white] (0,-0.25) circle (0.14);
    \node at (0.0 ,-0.75) {\small $s$};
    \node at (0.0 ,-0.25) {\small $\hat{s}$};
    \end{tikzpicture} = P
    \end{equation}
We can use this identity to our advantange.
In particular, we get that: 
\begin{equation}
\begin{tikzpicture} [scale = 1, baseline = {([yshift=-.5ex]current bounding box.center)}] 
    \draw[color = black] (-0.8, 0) -- (0.8, 0); 
    \draw[color = black] (-0.8, 0) -- (-0.8, 1);
    \draw[color = black] (0.8, 0) -- (0.8, 1);
    \draw[rounded corners, fill = lightorange(ryb)] (-1, 0.35) rectangle (1, 0.85);
    \node at (0, 0.58) {\small $V$};
    \draw[color = black] (0, 0) -- (0, -0.8); 
    \draw[fill = lightdodgerblue] (0,0) circle (0.24); 
    \node at (0.02,0) {\small $\mathcal{U}$};
    \draw[fill = white] (0,-0.5) circle (0.14);
    \node at (0.0,-0.5) {\small $s$};
\end{tikzpicture}\ \  =\ \ \begin{tikzpicture} [scale = 1, baseline = {([yshift=-.5ex]current bounding box.center)}] 
    \draw[color = black] (-0.8, 0) -- (0.8, 0); 
    \draw[color = black] (-0.8, 0) -- (-0.8, 1);
    \draw[color = black] (0.8, 0) -- (0.8, 1);
    \draw[rounded corners, fill = lightorange(ryb)] (-1, 0.35) rectangle (1, 0.85);
    \node at (0, 0.58) {\small $V$};
    \draw[color = black] (0, 0) -- (0, -1.6); 
    \draw[fill = lightdodgerblue] (0,0) circle (0.24); 
    \node at (0.02,0) {\small $\mathcal{U}$};
    \draw[fill = white] (0,-0.5) circle (0.14);
    \node at (0.0,-0.5) {\small $s$};
    \draw[fill = white] (0,-0.9) circle (0.14);
    \node at (0.1,-0.8) {\small $s^+$};
    \draw[fill = white] (0,-1.3) circle (0.14);
    \node at (0.0,-1.3) {\small $s$};
\end{tikzpicture}\ \  =\ \ 
    \begin{tikzpicture} [scale = 1, baseline = {([yshift=-.5ex]current bounding box.center)}] 
    \draw[color = black] (-0.8, 0) -- (0.8, 0); 
    \draw[color = black] (-0.8, 0) -- (-0.8, 1);
    \draw[color = black] (0.8, 0) -- (0.8, 1);
    \draw[rounded corners, fill = lightorange(ryb)] (-1, 0.35) rectangle (1, 0.85);
    \node at (0, 0.58) {\small $V$};
    \draw[color = black] (0, 0) -- (0, -1.2); 
    \draw[fill = lightdodgerblue] (0,0) circle (0.24); 
    \node at (0.02,0) {\small $\mathcal{U}$};
    \draw[fill = white] (0,-0.5) circle (0.14);
    \node at (0.0,-0.5) {\small $s$};
    \draw[fill = white] (0,-0.8) circle (0.14);
    \node at (0.075,-0.75) {\small $s^{\dagger}$};
    \draw[color = black] (-0.8, -1.3) -- (0.8, -1.3); 
    \draw[fill = lightdodgerblue] (0,-1.3) circle (0.24); 
    \node at (0.02,-1.3) {\small $\bar{\mathcal{U}}$};
    \draw[color = black] (-0.8, -1.3) -- (-0.8, -2.3);
    \draw[color = black] (0.8, -1.3) -- (0.8, -2.3);
     \draw[rounded corners, fill = lightorange(ryb)] (-1, -2.0) rectangle (1, -1.6);
    \node at (0, -1.8) {\small $V^{\dagger}$};
    \draw[color = black] (-0.8, -2.3) -- (0.8, -2.3); 
    \draw[color = black] (0, -2.3) -- (0, -3.5);
    \draw[fill = lightdodgerblue] (0,-2.3) circle (0.24); 
    \node at (0.02,-2.3) {\small $\mathcal{U}$}; 
    \draw[fill = white] (0,-0.5-2.3) circle (0.14);
    \node at (0.12,-0.47-2.3) {\small $(s^+)^{\dagger}$};
    \draw[fill = white] (0,-0.8-2.3) circle (0.14);
    \node at (0.1,-0.79-2.3) {\small $s^+$};
    \draw[color = black] (-0.8, -1.3-2.3) -- (0.8, -2.3 - 1.3); 
    \draw[fill = lightdodgerblue] (0,-2.3-1.3) circle (0.24); 
    \node at (0.02,-2.3-1.3) {\small $\bar{\mathcal{U}}$}; 
    \draw[color = black] (-0.8, -3.6) -- (-0.8, -3.6 -1);
    \draw[color = black] (0.8, -3.6) -- (0.8, -3.6 -1);
    \draw[color = black] (-0.8, -3.6 - 1) -- (0.8, -3.6 -1);
    \draw[color = black] (-0, -4.4) -- (-0, -5.4);
    \draw[fill = lightdodgerblue] (0,-2.3-1.3-1) circle (0.24); 
    \node at (0.02,-2.3-1.3-1) {\small ${\mathcal{U}}$}; 
    \draw[rounded corners, fill = lightorange(ryb)] (-1, -4.3) rectangle (1, -3.9);
    \node at (0.0,-4.1) {\small ${V}$}; 
    \draw[fill = white] (0,-5.1) circle (0.14);
    \node at (0.0,-5.1) {\small $s$};
    \end{tikzpicture}
    \ \ 
    = \ \ \begin{tikzpicture} [scale = 1, baseline = {([yshift=-.5ex]current bounding box.center)}] 
    \draw[color = black] (-0.8, 0) -- (0.8, 0); 
    %
    \draw[color = black] (0, 0) -- (0, -1.2); 
    \draw[fill = lightdodgerblue] (0,0) circle (0.24); 
    \node at (0.02,0) {\small $\mathcal{U}$};
    \draw[fill = white] (0,-0.5) circle (0.14);
    \node at (0.0,-0.5) {\small $s$};
    \draw[fill = white] (0,-0.8) circle (0.14);
    \node at (0.075,-0.75) {\small $s^{\dagger}$};
    \draw[color = black] (-0.8, -1.3) -- (0.8, -1.3); 
    \draw[fill = lightdodgerblue] (0,-1.3) circle (0.24); 
    \node at (0.02,-1.3) {\small $\bar{\mathcal{U}}$};
    \draw[color = black] (-0.8, -1.3) -- (-0.8, -2.3);
    \draw[color = black] (0.8, -1.3) -- (0.8, -2.3);
    \draw[color = black] (-0.8, -2.3) -- (0.8, -2.3); 
    \draw[color = black] (0, -2.3) -- (0, -3.5);
    \draw[fill = lightdodgerblue] (0,-2.3) circle (0.24); 
    \node at (0.02,-2.3) {\small $\mathcal{U}$}; 
    \draw[fill = white] (0,-0.5-2.3) circle (0.14);
    \node at (0.12,-0.47-2.3) {\small $(s^+)^{\dagger}$};
    \draw[fill = white] (0,-0.8-2.3) circle (0.14);
    \node at (0.1,-0.79-2.3) {\small $s^+$};
    \draw[color = black] (-0.8, -1.3-2.3) -- (0.8, -2.3 - 1.3); 
    \draw[fill = lightdodgerblue] (0,-2.3-1.3) circle (0.24); 
    \node at (0.02,-2.3-1.3) {\small $\bar{\mathcal{U}}$}; 
    \draw[color = black] (-0.8, -3.6) -- (-0.8, -3.6 -1);
    \draw[color = black] (0.8, -3.6) -- (0.8, -3.6 -1);
    \draw[color = black] (-0.8, -3.6 - 1) -- (0.8, -3.6 -1);
    \draw[color = black] (-0, -4.4) -- (-0, -5.4);
    \draw[fill = lightdodgerblue] (0,-2.3-1.3-1) circle (0.24); 
    \node at (0.02,-2.3-1.3-1) {\small ${\mathcal{U}}$}; 
    \draw[rounded corners, fill = lightorange(ryb)] (-1, -4.3) rectangle (1, -3.9);
    \node at (0.0,-4.1) {\small ${V}$}; 
    \draw[fill = white] (0,-5.1) circle (0.14);
    \node at (0.0,-5.1) {\small $s$};
    \end{tikzpicture}\ \ =\ \ \begin{tikzpicture} [scale = 1, baseline = {([yshift=-.5ex]current bounding box.center)}] 
    \draw[color = black] (-0.8, 0) -- (0.8, 0); 
    %
    \draw[color = black] (0, 0) -- (0, -1.2); 
    \draw[fill = lightdodgerblue] (0,0) circle (0.24); 
    \node at (0.02,0) {\small $\mathcal{U}$};
    \draw[fill = white] (0,-0.5) circle (0.14);
    \node at (0.0,-0.5) {\small $s$};
    \draw[fill = white] (0,-0.8) circle (0.14);
    \node at (0.1,-0.75) {\small $s^+$};
    \draw[color = black] (-0.8, -1.3) -- (0.8, -1.3); 
    \draw[fill = lightdodgerblue] (0,-1.3) circle (0.24); 
    \node at (0.02,-1.3) {\small $\bar{\mathcal{U}}$};
    \draw[color = black] (-0.8, -1.3) -- (-0.8, -2.3);
    \draw[color = black] (0.8, -1.3) -- (0.8, -2.3);
    \draw[color = black] (-0.8, -2.3) -- (0.8, -2.3);
    \draw[color = black] (0, -2.3) -- (0, -3.1);
    \draw[fill = white] (0,-2.8) circle (0.14);
    \node at (0.0,-2.8) {\small $s$};
    \draw[fill = lightdodgerblue] (0,-2.3) circle (0.24); 
    \node at (0.02,-2.3) {\small $\mathcal{U}$};
     \draw[rounded corners, fill = lightorange(ryb)] (-1, -2.0) rectangle (1, -1.6);
    \node at (0, -1.8) {\small $V$};
    \end{tikzpicture}
\end{equation}
where in the last step we used $s^+ (s^+)^{\dagger} s^{\dagger} s = s^+(s s^+)^{\dagger} s = s^+ s s^+ s = s^+ s$.

Note that since $(1 - P)$ annhilates the MPS, the above proves that:
\begin{equation}
    \begin{tikzpicture} [scale = 1, baseline = {([yshift=-.5ex]current bounding box.center)}] 
    \draw[color = black] (-0.8, 0) -- (0.8, 0); 
    \draw[color = black] (0, 0) -- (0, -0.8); 
    \draw[color = black] (-0.8, 0) -- (-0.8, 1);
    \draw[color = black] (0.8, 0) -- (0.8, 1);
    \draw[rounded corners, fill = lightorange(ryb)] (-1, 0.35) rectangle (1, 0.85);
    \draw[fill = lightdodgerblue] (0,0) circle (0.3); 
    \node at (-0.03,0) {\small $A$};
    \node at (0, 0.58) {\small $V$};
    \end{tikzpicture}
     =     \begin{tikzpicture} [scale = 1, baseline = {([yshift=-.5ex]current bounding box.center)}] 
    \draw[color = black] (-0.8, 0) -- (0.8, 0); 
    \draw[color = black] (0, 0) -- (0, -0.8); 
    \draw[fill = lightdodgerblue] (0,0) circle (0.3); 
    \node at (-0.03,0) {\small $A$};
    \node at (0.0, -1) {\small $U$};
    \end{tikzpicture}
\end{equation}
as desired.

\hspace{0.95\textwidth} $\blacksquare$

\section{Supplementary Materials B: Additional Details Regarding Preparable Phase Diagram}

In the main text, we provided a guided tour of the landscape of $\chi = 2$ preparable phases, which corresponded to points within a tetrahedron (with states related by permuting the axes being physically equivalent).
Here, we provide closed-form expressions for the trajectories shown in the text and provide a proof of the claim that the AKLT state is gluable with a physical deformation by an arbitrary on-site complex matrix.

\subsection{Physically Motivated Trajectories in the Tetrahedron}

In Fig.~\ref{fig:preparable_phase_diagram}, we showed several physically motivated trajectories of preparable quantum states in the preparable tetrahedron.
Specifically, we showed trajectories for the following families of states:
\begin{equation}
    \ket{\Psi_{1}(\beta)} = e^{\beta \sum_x X_x} \ket{\text{cluster}} \qquad \ket{\Psi_{2}(\beta)} = e^{\beta \sum_{x \text{ even}} X_x} \ket{\text{cluster}} \qquad \ket{\Psi_{3}(\beta)} = e^{\beta \sum_{n}(S^z_n)^2} \ket{\text{AKLT}} 
\end{equation}
where the first interpolated between the cluster state and the trivial state, the middle interpolated between the cluster state and the GHZ, and the final interpolated between the AKLT state and the N\'eel state.
The trajectories for each of these states are:
\begin{equation}
    \boldsymbol{\lambda}_1(\beta) = \frac{1}{2 (1 + \cosh(4 \beta))} \begin{pmatrix} e^{4 \beta}  \\ 1 \\ 1 \\ e^{- 4\beta}\end{pmatrix} \qquad \boldsymbol{\lambda}_2(\beta) = \frac{1}{4 \cosh(2\beta)} \begin{pmatrix} e^{2 \beta}  \\ e^{-2\beta} \\ e^{2\beta} \\ e^{- 2\beta}\end{pmatrix} \qquad \boldsymbol{\lambda}_3(\beta) = \frac{1}{1 + \cosh(2\beta)}\begin{pmatrix}
        0 \\ e^{2\beta} \\ e^{2\beta} \\ 1
    \end{pmatrix}
\end{equation}

\subsection{Arbitrary Invertible Complex Matrices on the AKLT State}

We now show that the AKLT state is preparable even if one acts at the physical level with an arbitrary complex matrix.
To do so, let us recall that the AKLT state is described by matrix product state tensors given by: 
\begin{equation}
\begin{tikzpicture}[scale = 1, baseline = {([yshift=-.5ex]current bounding box.center)}] 
        \draw[color = black] (-1, 0) -- (1, 0);
        \draw[color = black] (0, 0) -- (0, -0.8);
        \draw[fill = lightdodgerblue] (0,0) circle (0.3);
        \node at (-0.03,0) {\small $A$};
        \node at (0, -1) {\small $i$ };
    \end{tikzpicture} = \frac{1}{\sqrt{3}} \sigma^i
\end{equation}
in the appropriate basis \cite{tenpy} (the so-called Cartesian basis).
Furthermore, the AKLT tensor is symmetric under a full $SO(3)$ at the physical level, implying that any unitary acting at the physical level acts like a unitary gauge transformation of the MPS tensor.

Now, let us remark that, one can perform an SVD on $M$ and the result will be: 
\begin{equation}
    M = U S V^{\dagger} \qquad S = \text{diag} (s_1, s_2, s_3)
\end{equation}
where $s_1, s_2,$ and $s_3$ are in principle arbitrary but non-zero by assumption of invertibility.
Note that this means that any unitary matrix of the form: 
\begin{equation}
    W = U D U^{\dagger}
\end{equation}
where $D$ is diagonal will have the property that $WM = MW'$ for $W' = V^{\dagger} D V$.
Let us choose diagonal matrices to be $D_0 = \mathds{1}$, $D_1 = \text{diag}(1, -1, -1)$, $D_2 = \text{diag}(-1, 1, -1)$, $D_3 = (-1, -1, 1)$.
Note that $W_{\alpha} = U D_{\alpha} U^{\dagger}$ generates a $\mathbb{Z}_2 \times \mathbb{Z}_2$ subgroup of $SO(3)$ and hence have the property that:
\begin{equation}
\begin{tikzpicture}[scale = 1, baseline = {([yshift=-.5ex]current bounding box.center)}] 
        \draw[color = black] (-1, 0) -- (1, 0);
        \draw[color = black] (0, 0) -- (0, -0.8);
        \draw[fill = lightdodgerblue] (0,0) circle (0.3);
        \node at (-0.03,0) {\small $A$};
        \node at (0, -1) {\small $W_{\alpha}$ };
    \end{tikzpicture} =  \begin{tikzpicture}[scale = 1, baseline = {([yshift=-.5ex]current bounding box.center)}] 
        \draw[color = black] (-1, 0) -- (1, 0);
        \draw[color = black] (0, 0) -- (0, -0.8);
        \draw[fill = lightdodgerblue] (0,0) circle (0.3);
        \node at (-0.03,0) {\small $A$};
        \node at (-1.2, 0) {\small $V_{\alpha}$ };
        \node at (1.2, 0) {\small $V^{\dagger}_{\alpha}$ };
    \end{tikzpicture}
\end{equation}
where the $V_{\alpha}$'s form an irreducible faithful projective representation of $\mathbb{Z}_2 \times \mathbb{Z}_2$ with degree $2$ and hence forms a unitary error basis \cite{klappenecker2002beyond}.
The above discussion then implies that: 
\begin{equation}
\begin{tikzpicture}[scale = 1, baseline = {([yshift=-.5ex]current bounding box.center)}] 
        \draw[color = black] (-1, 0) -- (1, 0);
        \draw[color = black] (0, 0) -- (0, -0.8);
        \draw[fill = lightdodgerblue] (0,0) circle (0.3);
        \node at (-0.03,0) {\small $A$};
        \node at (-1.2, 0) {\small $V_{\alpha}$ };
        \draw[fill = lightorange(ryb)] (0, -0.5) circle (0.1);
    \end{tikzpicture} = 
\begin{tikzpicture}[scale = 1, baseline = {([yshift=-.5ex]current bounding box.center)}] 
        \draw[color = black] (-1, 0) -- (1, 0);
        \draw[color = black] (0, 0) -- (0, -0.8);
        \draw[fill = lightdodgerblue] (0,0) circle (0.3);
        \node at (-0.03,0) {\small $A$};
        \node at (0, -1) {\small $W'_{\alpha}$ };
        \node at (1.2, 0) {\small $V_{\alpha}$ };
        \draw[fill = lightorange(ryb)] (0, -0.5) circle (0.1);
    \end{tikzpicture} 
\end{equation}
for all $\alpha$, where the orange circle is the matrix $M$.
Consequently, the AKLT state deformed with an arbitrary invertible complex matrix is preparable.

\section{Supplementary Materials C: Preparable Matrix Product States for Arbitrary Nice Error Bases}

In the main text, we remarked that Eqs.~\eqref{eq:Achi2}~and~\eqref{eq:AchiN} could be generalized to arbitrary nice error bases, even non-abelian ones.
The generalization, expressed in terms of the transfer matrix, is given by: 
\begin{equation}
        \begin{tikzpicture} [scale = 1, baseline = {([yshift=-.5ex]current bounding box.center)}] 
    \draw[color = black] (-0.8, 0) -- (0.8, 0); 
    \draw[color = black] (-0.8, -1.2) -- (0.8, -1.2); 
    \draw[color = black] (0, 0) -- (0, -1.2); 
    \draw[fill = lightdodgerblue] (0,0) circle (0.3); 
    \draw[fill = lightdodgerblue] (0,-1.2) circle (0.3); 
    \node at (-0.03,0) {\small $A$};
    \node at (-0.03,-1.2) {\small $\bar{A}$};
    \end{tikzpicture} = \sum_{\mathcal{C} \text{ of } G_U} \lambda_{\mathcal{C}}\left(\frac{1}{|\mathcal{C}|} \sum_{g \in \mathcal{C}} \begin{tikzpicture} [scale = 1, baseline = {([yshift=-.5ex]current bounding box.center)}] 
    \draw[color = black] (-0.8, 0) -- (0.8, 0); 
    \draw[color = black] (-0.8, -1.2) -- (0.8, -1.2); 
    \draw[fill = white] (0,0) circle (0.25); 
    \draw[fill = white] (0,-1.2) circle (0.25); 
    \node at (0,0) {\small $V_g$};
    \node at (0,-1.2) {\small $\bar{V}_g$};
    \end{tikzpicture} \right)
    \end{equation}
where $\{V_g\}$ is a nice error basis, $G_U$ is its index group, and $\mathcal{C}$ label conjugacy classes of $G_U$.
By using the definition of nice error basis (i.e. $V_g V_h  =\omega(g,h) V_{gh}$ for some $2$-cocycle of $G_U$, $\omega$), one can readily show that the transfer matrix above satisfies the one-dimensional analog of Eq.~\eqref{eq:VTV=T}.
Namely, there is an error basis of $\chi^2$ operators that leave the transfer matrix invariant under conjugation.
This implies that the state can be prepared by using $\{V_g\}$ as a measurement basis.

\section{Supplementary Materials D: Generalization to Incomplete Basis Measurement}

In the main text, we briefly alluded to generalizations to incomplete basis measurement.
Here, we provide additional details for this.
The general idea will be to consider a tensor with the following structure:
\begin{equation}
    \begin{tikzpicture} [scale = 1.8, baseline = {([yshift=-.5ex]current bounding box.center)}] 
        \draw[color = black] (-1.5, 0) -- (1, 0);
        \draw[color = black] (-1, 0) -- (-1, -0.4);
        \draw[color = black] (-0.5, 0) -- (-0.5, -0.4);
        \draw[color = black] (0, 0) -- (0, -0.4);
        \draw[color = black] (0.5, 0) -- (0.5, -0.4);
        \draw[fill = lightdodgerblue] (-1, 0) circle (0.15);
        \draw[fill = lightdodgerblue] (-0, 0) circle (0.15);
        \node at (-1, 0) {\small $B$};
        \node at (0, 0) {\small $B$};
    \end{tikzpicture} 
\end{equation}
where the $T$-junctions are Kronecker deltas\footnote{Here, the Kronecker delta tensor is used for ease of demonstrating the point but generalizations of the idea should be clear.}
For simplicity, let us consider the case of $\chi = 2$.
Now, the only property that we require of $B$ is that: 
\begin{equation}
\begin{tikzpicture}[scale = 1, baseline = {([yshift=-.5ex]current bounding box.center)}] 
        \draw[color = black] (-1, 0) -- (1, 0);
        \draw[color = black] (0, 0) -- (0, -0.8);
        \draw[fill = lightdodgerblue] (0,0) circle (0.3);
        \node at (-0.03,0) {\small $B$};
        \node at (-1.2, 0) {\small $X$ };
    \end{tikzpicture}  = \begin{tikzpicture}[scale = 1, baseline = {([yshift=-.5ex]current bounding box.center)}] 
        \draw[color = black] (-1, 0) -- (1, 0);
        \draw[color = black] (0, 0) -- (0, -0.8);
        \draw[fill = lightdodgerblue] (0,0) circle (0.3);
        \node at (-0.03,0) {\small $B$};
        \node at (0, -1.0) {\small $U$};
        \node at (1.15, 0) {\small $X$};
    \end{tikzpicture}    
\end{equation}
In other words, only an incomplete basis of operators pushes through the $B$ tensor.
We can glue the above tensor network with incomplete basis measurement in the following way.
Let us first recognize a useful identity:
\begin{equation}
        \begin{tikzpicture}[scale = 1, baseline = {([yshift=-.5ex]current bounding box.center)}] 
        \draw[color = black] (0.7, -0.8) -- (-0.7 + 2.3, -0.8);
        \draw[fill = black] (1.6, -0.8) circle (0.07);
        \draw[color = black] (0.7, -0.4) -- (0.7, -1.2);
        \draw[color = black] (1.6, -0.4) -- (1.6, -1.2);
        \node at (0.7, -0.8) {$\small \bigoplus$};
        \node at (0.7, -1.4) {$\small \ket{0}$};
\end{tikzpicture}\ \  =\ \  
\begin{tikzpicture}[scale = 1, baseline = {([yshift=-.5ex]current bounding box.center)}]
    \draw[color = black] (-0.5, 0) -- (0.5, 0);
    \draw[color = black] (0, 0) -- (0, -0.5);
\end{tikzpicture}
\end{equation}
relating the control-NOT (CNOT) gate to the Kronecker delta tensor.
This naturally suggests an incomplete basis measurement protocol for the above tensor network.
In particular, first we prepare disentangled clusters given by the $B$ tensors.
Subsequently, instead of performing something like a Bell measurement between the clusters, which in practice would imply performing a CNOT gate between the virtual bonds and measuring the control and target in the $X$ and $Z$ bases respectively, we first entangle them with a CNOT and then only measure the target in the $Z$ basis.
The result will be: 
\begin{equation}
    \begin{tikzpicture}[scale = 1, baseline = {([yshift=-.5ex]current bounding box.center)}] 
        \foreach \i in {0, ..., 1}{
        \draw[color = black] (-0.7 + 2.3*\i, 0) -- (0.7 + 2.3*\i, 0);
        \draw[color = black] (-0.7 + 2.3*\i, 0) -- (-0.7 + 2.3*\i, -0.8);
        \draw[color = black] (0.7 + 2.3*\i, 0) -- (0.7 + 2.3*\i, -0.8);
        \draw[color = black] (0 + 2.3*\i, 0) -- (0 + 2.3*\i, -0.8);
        \draw[fill = lightdodgerblue] (0.0 + 2.3*\i, 0) circle (0.25);
        \node at (0.0 + 2.3*\i, 0) {\small $B$};
        }
        \draw[color = black] (0.7, -0.8) -- (-0.7 + 2.3, -0.8);
        \draw[fill = black] (1.6, -0.8) circle (0.07);
        \draw[color = black] (0.7, -0.8) -- (0.7, -1.2);
        \draw[color = black] (1.6, -0.8) -- (1.6, -1.2);
        \node at (0.7, -0.8) {$\small \bigoplus$};
        \node at (0.7, -1.4) {$\small \ket{0/1}$};
\end{tikzpicture}
\end{equation}
Note that in the case of measuring all $\ket{0}$ states, we have successfully prepared the matrix product state with incomplete basis measurements.
In the case, we measure $\ket{1}$, note that: 
\begin{equation}
    \begin{tikzpicture}[scale = 1, baseline = {([yshift=-.5ex]current bounding box.center)}] 
        \foreach \i in {0, ..., 1}{
        \draw[color = black] (-0.7 + 2.3*\i, 0) -- (0.7 + 2.3*\i, 0);
        \draw[color = black] (-0.7 + 2.3*\i, 0) -- (-0.7 + 2.3*\i, -0.8);
        \draw[color = black] (0.7 + 2.3*\i, 0) -- (0.7 + 2.3*\i, -0.8);
        \draw[color = black] (0 + 2.3*\i, 0) -- (0 + 2.3*\i, -0.8);
        \draw[fill = lightdodgerblue] (0.0 + 2.3*\i, 0) circle (0.25);
        \node at (0.0 + 2.3*\i, 0) {\small $B$};
        }
        \draw[color = black] (0.7, -0.8) -- (-0.7 + 2.3, -0.8);
        \draw[fill = black] (1.6, -0.8) circle (0.07);
        \draw[color = black] (0.7, -0.8) -- (0.7, -1.2);
        \draw[color = black] (1.6, -0.8) -- (1.6, -1.2);
        \node at (0.7, -0.8) {$\small \bigoplus$};
        \node at (0.7, -1.4) {$\small \ket{1}$};
\end{tikzpicture}\ \  =\ \ 
    \begin{tikzpicture}[scale = 1, baseline = {([yshift=-.5ex]current bounding box.center)}] 
        \foreach \i in {0, ..., 1}{
        \draw[color = black] (-0.7 + 2.3*\i, 0) -- (0.7 + 2.3*\i, 0);
        \draw[color = black] (-0.7 + 2.3*\i, 0) -- (-0.7 + 2.3*\i, -0.8);
        \draw[color = black] (0.7 + 2.3*\i, 0) -- (0.7 + 2.3*\i, -0.8);
        \draw[color = black] (0 + 2.3*\i, 0) -- (0 + 2.3*\i, -0.8);
        \draw[fill = lightdodgerblue] (0.0 + 2.3*\i, 0) circle (0.25);
        \node at (0.0 + 2.3*\i, 0) {\small $B$};
        }
        \draw[color = black] (0.7, -0.8) -- (-0.7 + 2.3, -0.8);
        \draw[fill = black] (1.6, -0.8) circle (0.07);
        \draw[color = black] (0.7, -0.8) -- (0.7, -1.2);
        \draw[color = black] (1.6, -0.8) -- (1.6, -1.2);
        \node at (0.7, -0.8) {$\small \bigoplus$};
        \node at (0.7, -1.4) {$\small \ket{0}$};
        \node at (0.7, -0.3) {$\small \textcolor{red}{X}$};
\end{tikzpicture}\ =\      \begin{tikzpicture}[scale = 1, baseline = {([yshift=-.5ex]current bounding box.center)}] 
        \draw[color = black] (-0.7 + 2.3*0, 0) -- (-0.7 + 2.3*1 + 2.3*0, 0);
        \draw[color = black] (-0.7 + 2.3*0, 0) -- (-0.7 + 2.3*0, -0.8);
        \draw[color = black] (0 + 2.3*0, 0) -- (0 + 2.3*0, -0.8);
        \draw[fill = lightdodgerblue] (0.0 + 2.3*0, 0) circle (0.25);
        \draw[color = black] (-0.7 + 2.3*1, 0) -- (0.7 + 2.3*1, 0);
        \node at (0.0 + 2.3*0, 0) {\small $B$};
        \draw[color = black] (0.7 + 2.3*1, 0) -- (0.7 + 2.3*1, -0.8);
        \draw[color = black] (0 + 2.3*1, 0) -- (0 + 2.3*1, -0.8);
        \draw[fill = lightdodgerblue] (0.0 + 2.3*1, 0) circle (0.25);
        \node at (0.0 + 2.3*1, 0) {\small $B$};
        \node at (1.0, 0) {$\small \textcolor{red}{X}$};
        \draw[color = black] (1.3, 0) -- (1.3, -0.8);
        \node at (0.7, -1.4) {$\small \ $};
\end{tikzpicture}
\end{equation}
which we can sweep through the chain and correct at the physical level.
Hence, states with the above form are deterministically correctable even with incomplete basis measurements.
As an example of a state with this structure, let us note that the MPS of the state $\ket{\Psi} = e^{\beta \sum_{x} Z_xZ_{x + 1}}\ket{+}^{\otimes N}$ takes the form: 
\begin{equation}
\ket{\Psi} = \begin{tikzpicture}[scale = 1, baseline = {([yshift=-.5ex]current bounding box.center)}] 
    \draw[color = black] (-3.5, 0) -- (3.5, 0);
    \foreach \i in {-3,...,3}{
        \draw[color = black] (\i, 0) -- (\i, -0.7);
    }
    \foreach \i in {-3,...,3}{
        \draw[fill = white] (\i + 0.3, 0) circle (0.1);
    }
    \end{tikzpicture}    
\end{equation}
where the white circles are $e^{\alpha X}$ with $\alpha = \text{arctanh}(e^{-2\beta})$.
Note that the above is of the desired form and has a ``$B$'' tensor with a trivial physical leg.
We recgonize the same structure in the toric code example in the main text.

\end{document}